# The CEDAR Workbench: An Ontology-Assisted Environment for Authoring Metadata that Describe Scientific Experiments

Rafael S. Gonçalves, Martin J. O'Connor, Marcos Martínez-Romero, Attila L. Egyedi, Debra Willrett, John Graybeal and Mark A. Musen

Stanford Center for Biomedical Informatics Research
Stanford University, CA, USA
`rafael.goncalves@stanford.edu`

**Abstract.** The Center for Expanded Data Annotation and Retrieval (CEDAR) aims to revolutionize the way that metadata describing scientific experiments are authored. The software we have developed—the CEDAR Workbench—is a suite of Web-based tools and REST APIs that allows users to construct metadata templates, to fill in templates to generate high-quality metadata, and to share and manage these resources. The CEDAR Workbench provides a versatile, REST-based environment for authoring metadata that are enriched with terms from ontologies. The metadata are available as JSON, JSON-LD, or RDF for easy integration in scientific applications and reusability on the Web. Users can leverage our APIs for validating and submitting metadata to external repositories. The CEDAR Workbench is freely available and open-source.

## 1   State of metadata in scientific repositories

There are vast amounts of scientific data hosted in a multitude of public repositories. These repositories are either discipline-specific, such as the Gene Expression Omnibus (GEO) for functional genomics data [1], or generic, such as the Zenodo repository for any type of data [2]. Despite the different types of content, these repositories share a common need for submitted data to be accompanied with precise, machine-interpretable descriptions of what the data represent—that is, *metadata*. Consider BioSample [3], a repository of metadata about samples used in biomedical experiments, maintained by the U.S. National Center for Biotechnology Information (NCBI). The data about these biological samples are typically associated with experimental data that are submitted elsewhere (e.g., GEO). For a better chance at understanding the associated experiment, or reusing the data to replicate that experiment, these resources should be appropriately linked by using agreed-upon terms, ideally from ontologies or other controlled term sources. A variety of studies have demonstrated that this linkage and rigorous typing rarely occur [4–6]. As a consequence, metadata in public repositories are typically weak. This lack of high-quality metadata hinders advancements in science, as the scientific community has difficulties reproducing findings or using existing data for new analyses [7]. To address this problem, the biomedical community developed dozens of metadata guidelines, which researchers can use to annotate experiment results. The so-called "minimal information" metadata guidelines specify the minimum information about experimental data that are necessary to ensure that the associated experiments can be reproduced. BioSharing [8]—a curated Web-based collection of data standards, databases, and policies in the life, environmental, and biomedical sciences—serves about a hundred of these "minimal information" guidelines and formats, such as the Minimal Information About a Microarray Experiment (MIAME) guideline [9]. However, such guidelines are typically loosely-defined and lack semantic linkage. For instance, GEO requests that investigators submit their datasets together with metadata that conform to the MIAME guideline. While MIAME specifies that submitters must include information for specified fields, it does not define how these values should be specified. As a result, typical GEO field values are unstructured free text. It is difficult to make an efficient use of these metadata when performing subsequent analyses.

The poor quality of metadata in scientific repositories is partly explained by the lack of appropriate tooling for producing high-quality metadata. Metadata repositories typically require spreadsheet-based submissions and specify a variety of *ad hoc* formats. To describe even a simple metadata submission using such formats demands significant effort on the author's part. Various tools exist to ease the burden of constructing metadata formats. The ISA Tools [10] provide a desktop application that allows users to construct spreadsheet-based submissions for metadata repositories, although there is no support for ontologies. The linkedISA software [11] adds mechanisms to annotate the spreadsheet-templates with controlled terms. Rightfield [12] is an Excel plugin that allows users to embed ontology-derived values in spreadsheets, and to restrict cell values to terms from ontologies. Annotare [13] is a desktop application similar to ISA Tools, although with support for using ontology terms. These tools all rely on spreadsheet-based representations, which are limited in their expressivity and difficult to extend. There is a need for software infrastructure based on an open format that is compliant with Web standards. The FAIR data principles [14] specify desirable criteria that metadata should meet. These data principles provide desiderata for a format and associated tooling for metadata authoring, which CEDAR [15] is developing.

## 2  CEDAR Workbench

With the goal of drastically improving the metadata that annotate datasets in public repositories, we built the CEDAR Workbench—a set of open-source, Web-based tools for the acquisition, storage, search, and reuse of metadata templates. The CEDAR Workbench offers its users the ability to construct metadata-acquisition forms or *templates*. The metadata produced using CEDAR templates are designed to be adherent to the FAIR data principles, and to be interoperable with Linked Open Data. CEDAR metadata is retrievable in JSON, JSON-LD, and RDF formats.
The CEDAR Workbench is used for generating metadata that describe scientific experiments. Users have access to metadata and associated metadata-authoring functionality using CEDAR's Web front-end or REST services. We host a public instance of the CEDAR Workbench at http://cedar.metadatacenter.net. The software is available on GitHub (http://github.com/metadatacenter) and released under the open-source 2-Clause BSD license. The project is described in full detail at http://metadatacenter.org.

### 2.1 System architecture

The CEDAR Workbench is a highly modular system, designed to allow its users to employ individual services in their applications or workflows. Fig. 1 shows the CEDAR Workbench architecture. The primary goal of CEDAR is to generate high-quality metadata describing scientific data that are semantically enriched with terms from ontologies. To that end, we developed a model that serves as a common, standards-based format for describing templates, fields, and metadata [16]. For interoperability on the Web, it is crucial that all resources be represented using an open model that can be serialized to widely accepted formats such as JSON-LD or any RDF syntax. We used JSON Schema and JSON-LD to encode the model. The model provides mechanisms for template composition to promote the reuse of templates.[1]

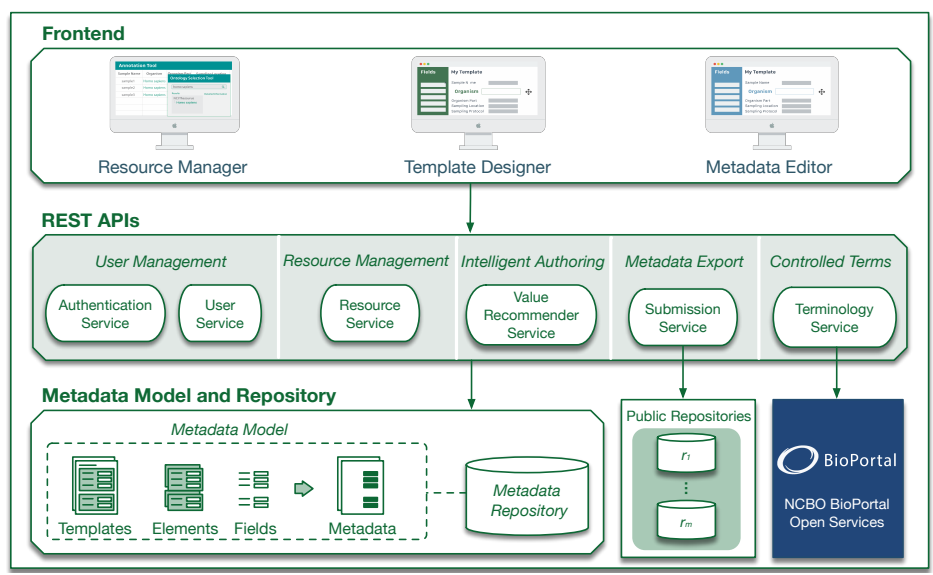

**Fig. 1.** Primary components of the CEDAR Workbench. Our software follows a microservice-based architecture. The system is built from a collection of loosely-couple services that provide self-contained functionality (e.g., User Service for user management). The CEDAR Workbench is composed of front-end components featuring a Resource Manager tool for managing and organizing resources into folders, a Template Designer for assembling templates, and a Metadata Editor for entering metadata. The Submission Service allows users to upload metadata to external, public repositories. The Terminology Service bridges CEDAR technology with BioPortal ontologies. All resources adhere to the CEDAR Metadata Model, and are stored in the Metadata Repository.

---
[1] Further details at: https://metadatacenter.org/tools-training/outreach/cedar-template-model.

The resources in the CEDAR Workbench—templates, fields, and metadata—are represented as JSON-LD documents that conform to our model. Templates and fields can be annotated with terms from ontologies [17] in the NCBO BioPortal—an online repository that serves as one of the primary platforms for hosting and sharing over 500 biomedical ontologies [18]. Because JSON-LD is a concrete syntax for RDF, all CEDAR metadata instances are RDF documents as well. All resources are stored in our Metadata Repository, which scientists can use to search for and browse templates in a faceted way.

The CEDAR Workbench microservices are implemented in Java using the Dropwizard framework (http://www.dropwizard.io), while the front-end is implemented in AngularJS (http://angularjs.org).

## 2.2 Main features

The overarching objective of the CEDAR Workbench is to make it easier and faster for users to annotate datasets with metadata. We target this goal by allowing users to build modular, customized metadata templates that can be filled out to create metadata.

**Resource Manager.** The Resource Manager is the primary front-end component. Using this tool, users can create templates and folders, search for metadata and templates, populate templates, and share resources.

**Template Designer.** In the Template Designer (Fig. 2), users can assemble metadata templates from other templates or fields. There are numerous field types and template formatting options available to template designers.

**BioPortal lookup service.** The CEDAR Workbench provides an interactive lookup service linked to BioPortal. This service allows template designers to find sets of ontology terms to annotate templates and fields—that is, to add type and property assertions using ontology classes and properties. Users can also specify that the possible values of fields must correspond to ontology terms. The classes and object, data, or annotation properties for performing these annotations can be selected from terms in BioPortal ontologies. When appropriate terms to do not exist, users can create new terms and value sets dynamically at template design-time (Fig. 2). Upon creating a new term, users can map it to existing terms in BioPortal ontologies using SKOS (http://www.w3.org/TR/skos-reference) properties [19].

**Intelligent Authoring.** To decrease metadata authoring time, we implemented a value recommendation feature that provides context-sensitive suggestions for input field values [20]. The value recommender learns associations between data values in metadata submissions, computes suggestions based on these associations, and presents the suggestions to metadata authors. The suggestions are ranked according to their applicability for each specific field. During metadata entry, users are prompted with drop-down lists and auto-complete suggestions given by the value recommender.

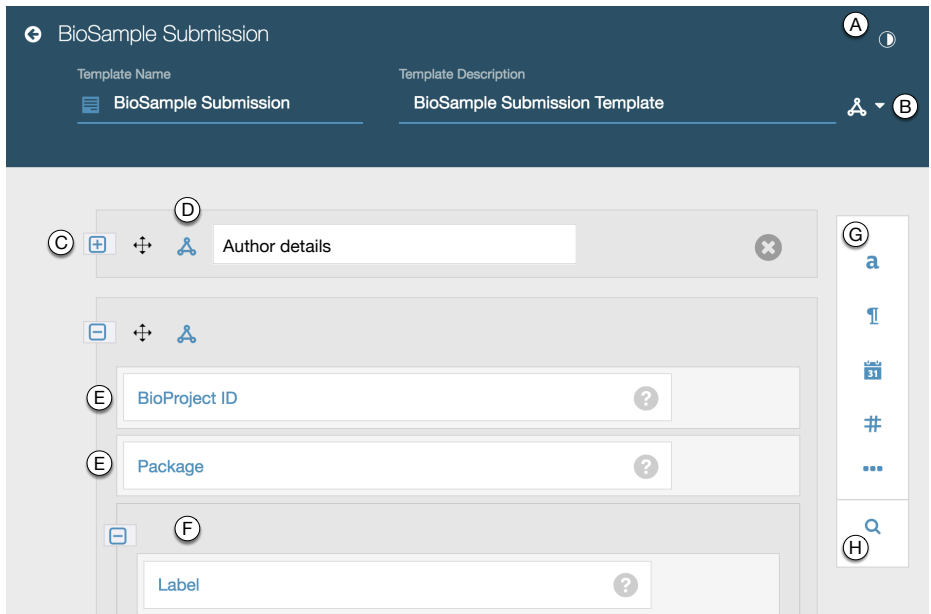

**Fig. 2.** Screenshot of the Template Designer loaded with a CEDAR template for a BioSample metadata submission. The annotated items in the screenshot are: (A) button to visualize the JSON Schema code corresponding to the template; (B) button to add terms from ontologies to annotate the template; (C) template element, which is collapsed and named; (D) button to add properties from ontologies to annotate the element; (E) fields that compose elements and templates; (F) element nested within an element; (G) field type options (in sight: text, paragraph, date, number, and a trigger for more options); (H) button to search for template elements to add.

**Metadata Editor.** The Metadata Editor is designed to facilitate rapid entry of metadata. This tool generates a streamlined, form-based acquisition interface based on a metadata template definition. Filling in metadata using the Metadata Editor is made easy with suggestions provided by the value recommender. When field values are constrained to a set of ontology terms, users select a term from the generated list of possible values with minimal effort.

**Validation.** With better metadata quality in mind, we designed validation features to improve the output of our tools. Metadata entered through CEDAR templates are automatically validated against the corresponding template's JSON Schema model to get immediate feedback regarding structural errors (e.g., a user enters a numeric value in a field where an ontology term is expected). Additionally, metadata can be validated against some existing, external validation service (such as a REST endpoint) that is provided as a parameter to CEDAR at template design-time. For example, the NCBI BioSample Validator [3], which validates the format and content of metadata submissions to BioSample, can be used with the BioSample template shown in Fig. 2.

**Collaboration.** The CEDAR Workbench provides a highly-collaborative environment, where users can create groups composed of their team members. Users can share all types of resources with individual users, among groups of users, or with the entire CEDAR community. When sharing resources, users can restrict access to these with common read/write permissions.

**REST API.** CEDAR REST APIs provide full access to the CEDAR ecosystem. Users can leverage the CEDAR API to export templates or metadata to other applications or repositories. The REST API is documented using Swagger and is described at https://metadatacenter.org/tools-training/cedar-api.

### 2.3    Community uptake

The CEDAR Workbench is used by several communities. The problems these communities face are primarily related to producing and managing FAIR metadata. In particular, the formats and tooling employed are based on spreadsheets that have no strict linkage to ontology terms. The Library of Integrated Network-Based Cellular Signatures (LINCS, http://www.lincsproject.org) is using CEDAR tools to build an end-to-end solution to submit data and metadata to the LINCS repository. The Human Immunology Project Consortium (HIPC, http://www.immuneprofiling.org) is implementing end-to-end workflows using the CEDAR Workbench to acquire and validate precisely-defined metadata entered by their users, which are then submitted to the BioSample or ImmPort repositories [21]. The Stanford Digital Repository (http://sdr.stanford.edu) in the Stanford University Libraries is testing the use of CEDAR templates for authoring metadata in several of their projects. These groups have encoded minimum information models as CEDAR templates, which they use in their submission pipelines. Note that none of these communities used ontology terms at all when authoring metadata. CEDAR helped to introduce semantics to the work that these groups carry out.

The AIRR community (http://airr-community.org) is developing standards for describing datasets acquired using sequencing technologies. The AIRR submission process involves submitting the generated metadata to the public NCBI BioSample repository. We built a submission pipeline to upload metadata to BioSample, which the AIRR community is successfully using. Based on our experience with the NCBI BioSample repository, we intend to generalize our submission infrastructure to other NCBI repositories.

These projects have succeeded in setting-up CEDAR-based metadata submission pipelines. The feedback from users in these communities is very positive—they find that working with CEDAR tools is straightforward, and that the metadata generated through CEDAR are of significantly higher quality than what they produced before. Our expectation is that progressively more communities will realize the potential that the CEDAR Workbench has for producing high-quality metadata.

### 3    Summary

We developed the CEDAR Workbench to improve the quality of metadata submitted to public repositories. CEDAR provides a freely-available suite of tools to build metadata templates, to fill them in with metadata, to submit the metadata to external repositories, and to store, search, and manage templates and metadata.

The novelty of our approach lies in the use of a principled, open format for the description of metadata resources, the ability to fill in metadata with guidance from intelligent authoring features, and finally, the ability to annotate templates and to restrict template field values to terms from ontologies. The CEDAR Workbench provides modular, highly-reusable components via a microservice-based architecture, allowing users to employ individual services for specific tasks, such as the BioPortal-linked Terminology Service. The metadata produced using CEDAR templates are FAIR-adherent by design, and are available in JSON, JSON-LD, and RDF formats for interoperability with Linked Open Data and Semantic Web applications.

Currently we use JSON Schema for imposing constraints on template input data. The SHACL candidate recommendation (http://www.w3.org/TR/shacl) may provide a more appropriate solution for constraining input data. However, SHACL is not yet standardized, and has limited tool support.

We are working toward allowing our users to submit metadata from CEDAR to an increasing number of external repositories. The submission pipelines we created are the first of many that will serve an increasing number of users, and help bring a semantic foundation to future metadata submission efforts.

**Acknowledgements**

CEDAR is supported by grant U54 AI117925 awarded by the National Institute of Allergy and Infectious Diseases through funds provided by the trans-NIH Big Data to Knowledge (BD2K) initiative (http://www.bd2k.nih.gov). NCBO is supported by the NIH Common Fund under grant U54HG004028.